\documentclass[twocolumn,showpacs,amsfonts,aps,prc,floatfix,%
superscriptaddress]{revtex4} 

\usepackage{amsmath}
\usepackage{bm}
\usepackage{graphicx}

\newcommand{\be}[1]{\begin{equation}\label{#1}}
\newcommand{\ee}{\end{equation}}

\newcommand{\lsim}{\mbox{\raisebox{-0.6ex}{$\stackrel{<}{\sim}$}}\:}

\usepackage{color}

\begin{document}


\title{Effects of Bulk Viscosity at Freezeout}
\date{\today}

\author{Akihiko Monnai}
\email{monnai@nt.phys.s.u-tokyo.ac.jp}
\affiliation{Department of Physics, The University of Tokyo,
Tokyo 113-0033, Japan}

\author{Tetsufumi Hirano}
\email{hirano@phys.s.u-tokyo.ac.jp}
\affiliation{Department of Physics, The University of Tokyo,
Tokyo 113-0033, Japan}

\begin{abstract}
We investigate particle spectra
and elliptic flow coefficients
in relativistic heavy ion collisions
by taking into account
the distortion of phase space distributions
due to bulk viscosity
at freezeout.
We first calculate the distortion of phase space
distributions in a multi-component system
within the Grad's fourteen moment method.
We find some subtle issues when one matches
macroscopic variables with microscopic
momentum distributions in a multi-component system
and develop a consistent
procedure to uniquely determine the corrections
in the phase space distributions.
Next, we calculate particle spectra
by using the Cooper-Frye formula
to see the effect of the bulk viscosity.
In spite of the relative smallness of the bulk viscosity,
we find that it is likely to have a visible effect
in particle spectra and elliptic flow coefficients.
This indicates the importance of bulk viscosity together
with shear viscosity if one wants to constrain
the transport coefficients with better accuracy from experimental data.
\end{abstract}
\pacs{25.75.-q, 25.75.Nq, 12.38.Mh, 12.38.Qk}

\maketitle

\section{Introduction}
\label{sec1}
\vspace*{-2mm}

One of the major discoveries at the Relativistic Heavy Ion Collider
(RHIC) in Brookhaven National Laboratory (BNL) \cite{experiments}
is that the quark-gluon plasma (QGP) behaves like a ``perfect liquid" \cite{BNL},
which attracts much theoretical interest in various fields.
Hydrodynamic modeling of heavy ion collisions
plays an important role in deducing the above fact \cite{reviews1,reviews2}.
Anisotropy of radial flow, namely elliptic flow \cite{Ollitrault},
is found to be large
in Au+Au and Cu+Cu collisions at RHIC energies \cite{STARv2,PHENIXv2,PHOBOSv2}.
Elliptic flow is quantified by the
second harmonics of azimuthal angle distribution,
$v_2 = \langle \cos 2 (\phi_p) \rangle$ \cite{PoskanzerVoloshin}.
One reproduces
the centrality, pseudorapidity, and transverse momentum
dependences of $v_2$ data reasonably well
by employing the Glauber type initial conditions and
implementing hadronic dissipative effects
in ideal hydrodynamic models \cite{HHKLN}.
However, when one replaces
initial conditions from the Glauber
model to the ones from a model based on the color glass condensate,
elliptic flow coefficients overshoot the experimental data
due to eccentricity larger
than that from the conventional Glauber model
calculations \cite{HHKLN}.
Given a fact that the initial conditions
in relativistic heavy ion collisions are not known precisely,
the discrepancy strongly suggests
effects of viscosities in the QGP phase are required to 
reproduce the $v_2$ data in this particular case.
Even though ideal hydrodynamic models have been successful in 
reproduction of the $v_2$ data,
the next important tasks are to constrain the transport coefficients
by comparison of theoretical predictions with the experimental data
and to comprehensively understand the transport properties
of the QGP. 

Shear viscosity has already been taken into account
in several hydrodynamic simulations
to investigate its effect on elliptic flow 
in relativistic heavy ion collisions \cite{Song:2007fn,Song:2007ux,Song:2008si,Chaudhuri,Baier,Dusling}.
On the other hand, few studies
are done so far
to investigate the effect of bulk viscosity
 in dynamical simulations
of relativistic viscous fluids \cite{Paech:2006st,Koide,Torrieri:2008ip,Fries,Song:2008hj}.
Many years ago, it was found that the bulk viscous
coefficient $\zeta$ has a prominent peak 
in the vicinity of the quantum chromodynamics (QCD) phase transition
due to reduction of the sound velocity \cite{Mizutani:1987wb}.
Recently, the importance of bulk viscosity is 
realized again in the context of violation of scale invariance in QCD \cite{Kharzeev:2007wb}.
The smallness of the sound velocity is intimately related
with violation of scale invariance.
So it would be a universal feature
that the bulk viscous coefficient
becomes large around the critical region.
From a phenomenological point of view,
large bulk viscosity
could trigger a violation of applicability of hydrodynamic framework
around (pseudo-)critical temperature \cite{Torrieri:2007fb}
and support the success of hybrid approaches \cite{HHKLN, Teaney:2000cw, Nonaka:2006yn} in which
the hydrodynamic description of the QGP is followed by
the kinetic description of the hadron gas.

In this paper,
we focus on the effects of bulk viscosity
in the late stage of relativistic heavy ion 
collisions. 
To demonstrate them,
we calculate transverse momentum spectra and
differential elliptic flow coefficient.
In ideal hydrodynamic calculations, the Cooper-Frye formula \cite{Cooper:1974mv} is
conventionally employed to calculate the particle spectra at freezeout.
In the hybrid calculations, it plays
a role in transition of description
from macroscopic hydrodynamics to microscopic
hadronic cascade models.
In both cases, the Cooper-Frye formula
is indispensable in comparison of hydrodynamic results
with experimental data.
In dissipative hydrodynamics, viscous effects
are taken into account in the Cooper-Frye formula in two ways:
One is a variation of flow profile
as a matter of viscous correction
to the dynamical evolution,  the other is
distortion of phase space distribution
from its equilibrium form.
So far, no full three dimensional
viscous hydrodynamic simulations are available.
Given this situation, 
we take 
flow profile from a full (3+1)-dimensional ideal hydrodynamic simulation \cite{HHKLN, Hiranov2eta}
assuming variation of it due to viscosity is not significant.
We estimate the correction from bulk viscosity as well as from shear viscosity
to the distribution function
by matching macroscopic quantities with the ones
calculated in the kinetic theory
and see how this affects transverse momentum ($p_T$) spectra
and transverse momentum dependence of elliptic flow coefficient $v_2(p_T)$.

This paper is organized as follows.
In Sec.~\ref{sec2}, we give
a brief overview on the relativistic kinetic theory
by putting emphasizes on subtle issues
in its application to relativistic multi-component gases
and to the ones with zero (net-baryon) number density limit.
In Sec.~\ref{sec3},
we focus on a hadron resonance gas model, which is
widely exploited in hydrodynamic analyses
of relativistic heavy ion collisions,
and introduce a simple model for shear and bulk
viscous coefficients. 
We demonstrate the effects of bulk viscosity at freezeout
is possibly large by calculating the prefactors
in correction of distribution functions.
In Sec.~\ref{sec4}, we calculate $p_T$-spectra and $v_2(p_T)$
by taking flow profiles from three cases:
(1) Bjorken flow, (2) a blast wave model,
and (3) a full (3+1)-dimensional ideal hydrodynamic simulation.
Section \ref{sec5} is devoted to conclusion.


\section{The relativistic kinetic theory}
\label{sec2}

We would like to express viscous corrections to the phase space distribution for a relativistic multi-component gas in terms of macroscopic variables by matching the macroscopic variables with the ones obtained from the kinetic theory. This is a generalization of the treatment developed by Israel and Stewart \cite{Israel:1979wp} for a single component system to a multi-component system. 

We will see that some non-trivial facts arise when a multi-component gas is considered. Firstly, one usually considers a scalar term, a vector term, and a traceless tensor term for a momentum expansion of the non-equilibrium part of the distribution and assumes that the trace part of the tensor term can be absorbed in the scalar term. However, this is no longer the case for a multi-component system, because the equivalence of the scalar term and the trace part holds only in a single component gas. Secondly, the number of equations which determines the modification of the distribution does not change upon taking a zero net baryon density limit. This means we do not have to suffer the loss of conditions to obtain the unique expression of the distortion, nor do we need to resort to the quadratic ansatz \cite{Dusling}. Consideration of the limit is of practical importance since the limit is often employed in hydrodynamic models of the hot QCD matter created in relativistic heavy ion collisions.

\subsection{Distortion of the Distribution Function}

Tensor decompositions of the energy-momentum tensor $T^{\mu \nu}$ and the net baryon number current $N^\mu_B$ give the definitions of the macroscopic variables. The decomposed terms are put into the equilibrium part and the non-equilibrium part:
\begin{eqnarray}
\label{eq:em_eqnoneq}
T ^{\mu \nu} &=& T ^{\mu \nu}_0 + \delta T ^{\mu \nu} ,
\end{eqnarray}
where
\begin{eqnarray}
\label{eq:em_decomposition}
T_0 ^{\mu \nu} &= & e_0 u^\mu u^\nu - P \Delta ^{\mu \nu}, \\
\delta T ^{\mu \nu} &= & - \Pi \Delta ^{\mu \nu} + 2W^{( \mu} u^{\nu )} + \pi^{\mu \nu},
\end{eqnarray}
and
\begin{eqnarray}
\label{eq:n_eqnoneq}
N_B ^{\mu} &=& N_{B0}^{\mu} + \delta N_B^{\mu},
\end{eqnarray}
where
\begin{eqnarray}
\label{eq:n_decomposition}
N_{B0}^{\mu} &= & n_{B0} u^{\mu}, \ \delta N_B^{\mu} = V^{\mu}.
\end{eqnarray}
The time-like and space-like 
projection operators are, respectively,
defined as $u^\mu$ and $\Delta^{\mu \nu} = g^{\mu \nu}-u^\mu u^\nu$ with the Minkowski metric $g^{\mu \nu} = \mathrm{diag}(+,-,-,-)$. We also define the dot product as $a\cdot b = a_\mu b_\nu g^{\mu \nu}$. The round brackets stand for symmetrization,  $A^{( \mu} B^{\nu )}= \frac{1}{2}(A^\mu B^\nu + A^\nu B^\mu)$. 
$e_0 = u_\mu T_0^{\mu \nu} u_\nu$, $P = -\frac{1}{3}  \Delta_{\mu \nu} T_0^{\mu \nu}$, and $n_{B0} = u_\mu N_{B0}^\mu$ denote the energy density, the hydrostatic pressure, and the charge density, respectively.
$W^\mu = \Delta ^\mu _{\ \alpha} \delta T^{\alpha \beta} u_\beta $ is the energy current, $V^\mu = \Delta ^\mu _{\ \nu} \delta N^\nu_B$ the charge current, $\Pi = -\frac{1}{3} \Delta_{\mu \nu} \delta T^{\mu \nu}$ the bulk pressure, and $\pi ^{\mu \nu} = T ^{\langle \mu \nu \rangle} = [ \frac{1}{2} (\Delta ^\mu _{\ \alpha} \Delta ^\nu _{\ \beta} + \Delta ^\mu _{\ \beta} \Delta ^\nu _{\ \alpha}) - \frac{1}{3}\Delta^{\mu \nu} \Delta_{\alpha \beta}] T^{\alpha \beta} $ the shear stress tensor. Generally $e_0$ and $n_{B0}$ would have dissipative counterparts $\delta e$ and $\delta n_B$, but such terms are not considered here because we employ the Landau matching conditions which demand $\delta e = u_\mu \delta T^{\mu \nu} u_\nu = 0$ and $\delta n_B = u_\mu \delta N^{\mu}_B = 0$. These conditions are necessary to make the system thermodynamically stable in the first order theory. We will see the details later. For reviews on relativistic viscous hydrodynamics, see Ref.~\cite{reviews2}.

Kinetic theory demands for a system with multi-species particles
\begin{eqnarray}
\label{eq:energy-momentum}
T ^{\mu \nu} &=& \sum _i \int \frac{g_i d^3 p}{(2\pi )^3 E_i} p_i^\mu p_i^\nu f_i, \\
\label{eq:number}
N ^{\mu }_{B} &=& \sum _i \int \frac{b_i g_i d^3 p}{(2\pi )^3 E_i} p_i^\mu f_i .
\end{eqnarray}
The label $i$ denotes a particle species. $g_i$ and $b_i$ are the degeneracy and the baryon number. We define $\delta f_i = f_i - f_i^0$ to be the deviation of the distribution function from its equilibrium form, $f_i^0 = [\exp{(\frac{p_i \cdot u - b_i \mu_B}{T})} - \epsilon ]^{-1} = (\exp{y_0} - \epsilon )^{-1}$. Here $\epsilon = +1$ for bosons and $\epsilon = -1$ for fermions. $T$ and $\mu _B$ denote the temperature and the baryon chemical potential.

Now we expand $y$ defined by $f_i = (\exp{y} - \epsilon )^{-1}$ up to the second order in momentum and estimate viscous corrections to the distribution function in the Grad's 14-moment method
\begin{eqnarray}
\label{eq:df}
\delta f^i &=& - f_0^i (1+\epsilon f_0^i) (p_i^\mu \varepsilon _{\mu} + p_i^\mu p_i^\nu \varepsilon _{\mu \nu} ) ,
\end{eqnarray}
where $\varepsilon_{\mu}$ and $\varepsilon_{\mu \nu}$ are coefficients of the expansion. The numbers of unknown variables are 4 and 10, respectively,  because the latter should be symmetric since any anti-symmetric term cancels out upon taking contraction with $p_i^\mu p_i^\nu$.

$\varepsilon_{\mu \nu}$ is often considered to be traceless and the scalar term $\varepsilon$ is introduced instead \cite{Israel:1979wp}. The apparent lack of the scalar term in the expression (\ref{eq:df}) is due to the fact that we set the tensor term $\varepsilon _{\mu \nu}$ to have a non-zero trace. The numbers of unknown variables are the same in both cases. If we consider a single component system such as a pion gas, the trace part of the tensor term is equivalent to a scalar term $\varepsilon$ since the trace part can be separated as $\frac{\mathrm{Tr} (\varepsilon _{\mu \nu})}{4} g_{\mu \nu}$ and the metric produces a scalar $p^\mu p^\nu g_{\mu \nu}=m^2$.

However, this is not the case for a multi-component system because all the viscous correction tensors $\varepsilon$, $\varepsilon _{\mu}$, and $\varepsilon _{\mu \nu}$ are macroscopic quantities in the sense that
they do not depend on any particular particle species.
On the other hand, the trace part has mass dependence, which means it is particle-species dependent:
\begin{eqnarray}
\delta f^i _\mathrm{tensor} &=& - f_0^i (1+\epsilon f_0^i) p_i^\mu p_i^\nu \varepsilon _{\mu \nu} 
\notag \\
&=& - f_0^i (1+\epsilon f_0^i) p_i^\mu p_i^\nu \bigg[ \frac{\mathrm{Tr} (\varepsilon _{\mu \nu})}{4} g_{\mu \nu} \notag \\
&+& \bigg( \varepsilon _{\mu \nu} - \frac{\mathrm{Tr} (\varepsilon _{\mu \nu})}{4} g_{\mu \nu} \bigg) \bigg]
\notag \\
\label{eq:scalar_trace}
&= & - f_0^i (1+\epsilon f_0^i) \bigg[ \frac{\mathrm{Tr} (\varepsilon _{\mu \nu}) m_i^2 }{4} + p_i^\mu p_i^\nu \tilde{\varepsilon} _{\mu \nu} \bigg] .
\end{eqnarray}
Here $\tilde{\varepsilon} _{\mu \nu}$ is a traceless tensor. The equivalence of the scalar term and the trace term holds for a single component system because we can set a variable $\mathrm{Tr}(\varepsilon_{\mu \nu})$ to be $\frac{4\varepsilon }{m^2}$. Since the same prescription does not work for a multi-component gas where we have the additional index $i$, it is problematic whether to consider a non-zero trace tensor term, or the combination of a scalar term and a traceless tensor term in a system of multi-species particles. We will see in the next section that only the former is relevant in the case of a 16-component hadron resonance gas.

Our aim here is to uniquely determine the explicit forms of $\varepsilon_{\mu}$ and $\varepsilon_{\mu \nu}$. As we mentioned earlier, we introduce the Landau matching conditions $u_\mu \delta T^{\mu \nu} u_\nu = 0$ and $u_\mu \delta N^{\mu}_B = 0$, which ensure the thermodynamic stability in the first order theory (see Appendix \ref{lmc} for further details). They demand the energy density and the number density to match the equilibrium densities. Together with the kinetic definitions of the macroscopic variables, we now have 14 equations in total to determine 14 unknowns in $\delta f$. 

We first define $J^{\mu _1 \mu _2 ... \mu _m}$ and $\tilde{J}^{\mu _1 \mu _2 ... \mu _m}$ to express the conditions explicitly in terms of thermodynamic quantities
\begin{eqnarray}
\label{eq:jmn}
J^{\mu _1 \mu _2 ... \mu _m} &= & \sum _i \int \frac{g_i d^3 p}{(2\pi )^3 E_i} f_0^i (1+\epsilon f_0^i) p_i^{\mu _1} p_i^{\mu _2} ... p_i^{\mu _m} \notag \\
&= & \sum _n \big[ (\Delta ^{\mu _1 \mu _2} ... \Delta ^{\mu _{2n-1} \mu _{2n}} u^{\mu _{2n+1}} ... u^{\mu _m}) \notag \\
&+ & \mathrm{(permutations)} \big] J_{mn}, \\
\label{eq:tildejmn}
\tilde{J}^{\mu _1 \mu _2 ... \mu _m} &= & \sum _i \int \frac{b_i g_i d^3 p}{(2\pi )^3 E_i} f_0^i (1+\epsilon f_0^i) p_i^{\mu _1} p_i^{\mu _2} ... p_i^{\mu _m} \notag \\
&= & \sum _n \big[ (\Delta ^{\mu _1 \mu _2} ... \Delta ^{\mu _{2n-1} \mu _{2n}} u^{\mu _{2n+1}} ... u^{\mu _m}) \notag \\
&+ & \mathrm{(permutations)} \big] \tilde{J}_{mn}.
\end{eqnarray}
$J_{mn}$ is a coefficient of the expansion of $J^{\mu _1 \mu _2 ... \mu _m}$ by $(m-2n)$ $u^\mu$'s and $n$ $\Delta ^{\mu \nu}$'s. $\tilde{J}_{mn}$ is defined in the same way, but has the baryon number $b_i$ as an weight factor. These quantities should be distinguished in a mixture system of baryons, anti-baryons, and mesons because they contribute differently to the energy-momentum tensor and the baryon number current as seen in Eqs. (\ref{eq:energy-momentum}) and (\ref{eq:number}).

The Landau matching condition for the energy-momentum tensor is simplified by using the expressions defined above:
\begin{eqnarray}
\label{eq:lce1}
0 &=& u_\mu u_\nu \sum _i \int \frac{g_i d^3 p}{(2\pi )^3 E_i} p_i^\mu p_i^\nu \delta f_i \notag \\
&=& u_\mu u_\nu J^{\mu \nu \alpha} \varepsilon _\alpha + u_\mu u_\nu J^{\mu \nu \alpha \beta} \varepsilon _{\alpha \beta} \notag \\
&=& -J_{30} \varepsilon _{*} -(J_{40}-J_{41}) \varepsilon _{**} - J_{41} \mathrm{Tr}(\varepsilon ).
\end{eqnarray}
From now on we employ the notations $\varepsilon _{*} = \varepsilon _\mu u^\mu$, $\varepsilon _{**} =  \varepsilon _{\mu \nu} u^\mu u^\nu$, $\mathrm{Tr}(\varepsilon ) =  \varepsilon _{\mu}^{\mu}$, and $\Delta ^{\mu \nu} \varepsilon _{\nu *} = \Delta ^{\mu \nu} u^\alpha \varepsilon _{\nu \alpha}$. The other conditions can be expressed in a similar fashion:
\begin{eqnarray}
\label{eq:lcn1}
0 &=& -\tilde{J}_{20} \varepsilon _{*} -(\tilde{J}_{30}-\tilde{J}_{31}) \varepsilon _{**} - \tilde{J}_{31} \mathrm{Tr}(\varepsilon ), \\
\label{eq:bc1}
\Pi &=& J_{31} \varepsilon _{*} + \bigg( J_{41}-\frac{5}{3}J_{42} \bigg) \varepsilon _{**} + \frac{5}{3} J_{42} \mathrm{Tr}(\varepsilon ), \\
\label{eq:hc1}
W^\mu &=& - J_{31} \Delta ^{\mu \nu} \varepsilon _{\nu} - 2 J_{41} \Delta ^{\mu \nu} \varepsilon _{\nu *}, \\
\label{eq:cc1}
V^\mu &=& - \tilde{J}_{21} \Delta ^{\mu \nu} \varepsilon _{\nu} - 2 \tilde{J}_{31} \Delta ^{\mu \nu} \varepsilon _{\nu *}, \\
\label{eq:sc1}
\pi ^{\mu \nu} &=& -2 J_{42} \varepsilon ^{\langle \mu \nu \rangle}.
\end{eqnarray}
The conditions (\ref{eq:lce1})-(\ref{eq:sc1}) are classified into three independent sets of equations: (a) the definition of the bulk pressure and the Landau matching conditions for the scalars $\varepsilon _{*}$, $\varepsilon _{**}$, and $\mathrm{Tr}(\varepsilon )$, (b) the definitions of the energy current and the charge current for the vectors $\Delta ^{\mu \nu} \varepsilon _{\nu}$ and $\Delta ^{\mu \nu} \varepsilon _{\nu *}$, and (c) the definition of the shear stress tensor for the tensor $\varepsilon ^{\langle \mu \nu \rangle}$. The equations are immediately solved for each group:
\begin{eqnarray}
\varepsilon _{*} &=& \varepsilon _\mu u^\mu = D_0 \Pi , \label{D0s} \\
\varepsilon _{**} &=& \varepsilon _{\mu \nu} u^\mu u^\nu = \tilde{B}_0 \Pi , \label{B0s} \\
\mathrm{Tr}(\varepsilon ) &=& \varepsilon _{\mu}^{\mu} = B_3 \Pi , \label{B3s} \\
\Delta ^{\mu \nu} \varepsilon _{\nu} &=& D_1 W^\mu + \tilde{D}_1 V^\mu , \label{D1s} \\
\Delta ^{\mu \nu} \varepsilon _{\nu *} &=& B_1 W^\mu + \tilde{B}_1 V^\mu, \label{B1s} \\
\varepsilon ^{\langle \mu \nu \rangle} &=& B_2 \pi ^{\mu \nu}, \label{B2s}
\end{eqnarray}
where the prefactors $D_i$ and $B_i$ in Eqs. (\ref{D0s})-(\ref{B2s}) are functions of $J_{mn}$ and $\tilde{J}_{mn}$. Thus $\varepsilon_{\mu}$ and $\varepsilon_{\mu \nu}$ are determined to be
\begin{eqnarray}
\label{eq:linear}
\varepsilon _{\mu} &=& \varepsilon _{*} u_\mu + \Delta _{\mu \nu} \varepsilon ^{\nu} \notag \\
&=& D_0 \Pi u_\mu + D_1 W_\mu + \tilde{D}_1 V_\mu , \\
\varepsilon _{\mu \nu} &=& \varepsilon^{**} u_\mu u_\nu + \Delta_{\mu \nu} (\mathrm{Tr}(\varepsilon)- \varepsilon^{**})/3 \notag \\
&+& 2 u_{(\mu} \Delta_{\nu ) \alpha} \varepsilon ^{\alpha *} + \varepsilon_{\langle \mu \nu \rangle} \notag \\
&=& (B_0 \Delta _{\mu \nu} + \tilde{B}_0 u_\mu u_\nu )\Pi \notag \\
\label{eq:quadratic}
&+& 2 B_1 u_{( \mu} W_{\nu )} + 2 \tilde{B}_1 u_{( \mu} V_{\nu )} + B_2\pi_{\mu \nu},
\end{eqnarray}
where $B_0 = (B_3 - \tilde{B}_0)/3$ is employed. The prefactors are explicitly expressed as
\begin{eqnarray}
\label{eq:D0}
D_0 &=& 3 (J_{40} \tilde{J}_{31} - J_{41} \tilde{J}_{30}) \mathcal{J}_3^{-1} ,\\
\label{eq:B0}
B_0 &=& (J_{30} \tilde{J}_{30} - J_{40} \tilde{J}_{20}) \mathcal{J}_3^{-1} ,\\
\label{eq:B0tilde}
\tilde{B}_0 &=& 3 (J_{41} \tilde{J}_{20} - J_{30} \tilde{J}_{31}) \mathcal{J}_3^{-1} ,\\
\label{eq:D1}
D_1 &=& -2 \tilde{J}_{31} \mathcal{J}_2^{-1} ,\\
\label{eq:D1tilde}
\tilde{D}_1 &=& 2 J_{41} \mathcal{J}_2^{-1} ,\\
\label{eq:B1}
B_1 &=& \tilde{J}_{21} \mathcal{J}_2^{-1} ,\\
\label{eq:B1tilde}
\tilde{B}_1 &=& - J_{31} \mathcal{J}_2^{-1} ,\\
\label{eq:B2}
B_2 &=& \mathcal{J}_1^{-1} ,
\end{eqnarray}
where
\begin{eqnarray}
\mathcal{J}_3 &= & 5 J_{30} J_{42} \tilde{J}_{30} + 3 J_{31} J_{40} \tilde{J}_{31} + 3 J_{41} J_{41} \tilde{J}_{20} \notag \\
\label{eq:detJ3}
&-& 3 J_{31} J_{41} \tilde{J}_{30} - 3 J_{30} J_{41} \tilde{J}_{31} - 5 J_{40} J_{42} \tilde{J}_{20}, \\
\label{eq:detJ2}
\mathcal{J}_2 &= & 2 J_{31}\tilde{J}_{31} - 2 J_{41}\tilde{J}_{21}, \\
\label{eq:detJ1}
\mathcal{J}_1 &= & -2 J_{42} .
\end{eqnarray}
The prefactor for the shear stress tensor (\ref{eq:B2}) has the same form as shown in Ref.~\cite{Israel:1979wp}, whereas those for the bulk pressure (\ref{eq:D0})-(\ref{eq:B0tilde}) are different from the ones in the reference because the finite trace tensor term is considered instead of the scalar term.
Of course, those reduce to the same formulas in a single component gas.

It is worthwhile to mention that, if we define the heat current $q^\mu$ as
\begin{eqnarray}
\label{eq:q}
q_\mu = W_\mu - \frac{J_{31}}{\tilde{J}_{21}} V_\mu = W_\mu - \frac{e+P}{n_B} V_\mu,
\end{eqnarray}
Eqs.~(\ref{D1s}) and (\ref{B1s}) can be expressed as $D_1 q_\mu - V_\mu /n_BT$ and $B_1 q_\mu$, respectively.

\subsection{Discussion}

Some comments are in order here.

Firstly, we give a consideration to the special case where the Landau frame, \textit{i.e.}, $W^\mu =0 $ and the zero net baryon density limit $N_B^\mu \to 0$ are employed. These are often taken for hydrodynamic analyses of the QGP in relativistic heavy ion collisions because the numbers of baryons and anti-baryons are roughly the same for such events. In our framework, they correspond to estimating only the contributions of the shear viscosity and the bulk viscosity.

Apparently the Landau condition of the charge current (\ref{eq:lcn1}) vanishes upon taking the limit because $\tilde{J}_{mn}$ does. This could induce an ambiguity because the number of conditions for the determination of the prefactors decreases. But if we expand $\tilde{J}_{mn}$ around the baryon chemical potential $\mu _B = 0$, we have
\begin{equation}
\label{eq:jmntilde}
\tilde{J}_{mn} = 0 + \mu _B \frac{\partial \tilde{J}_{mn}}{\partial \mu _B} \bigg| _{\mu _B = 0} + \mathcal{O}(\mu _B ^2).
\end{equation}
Then Eq. (\ref{eq:lcn1}) is reduced to 
\begin{equation}
\label{eq:nochemlimit}
\bigg[ \frac{\partial \tilde{J}_{20}}{\partial \mu _B} \varepsilon _{*} + \frac{\partial (\tilde{J}_{30} -  \tilde{J}_{31})}{\partial \mu _B} \varepsilon _{**} + \frac{\partial \tilde{J}_{31}}{\partial \mu _B} \mathrm{Tr}(\varepsilon ) \bigg] _{\mu _B = 0} = 0,
\end{equation}
which remains finite in the zero baryon density limit. This enables us to determine the prefactors for the bulk pressure uniquely even after the limit is taken.

Secondly, we discuss the validity of the quadratic ansatz
in the distortion of the distribution.
If one makes the quadratic ansatz, \textit{i.e.}, $\varepsilon _{\mu \nu} = C_1 \pi _{\mu \nu} +C_2 \Delta _{\mu \nu} \Pi  $ \cite{Dusling}, then it violates the Landau conditions because
\begin{eqnarray}
\label{eq:nolce}
u_\mu \delta T^{\mu \nu} u_\nu &=& -J_{41} \mathrm{Tr}(\varepsilon ) \neq 0, \\
\label{eq:nolcn}
u_\mu \delta N^{\mu}_B &=& -\tilde{J}_{31} \mathrm{Tr}(\varepsilon ) \neq 0.
\end{eqnarray}
Therefore, one should not make the ansatz if one employs the Landau matching
conditions in hydrodynamic equations.
Note that, in the quadratic ansatz, $\varepsilon_{**}$'s in Eqs. (\ref{eq:lce1}) and (\ref{eq:lcn1}) vanish since $\varepsilon _{\mu \nu}$ is perpendicular to $u^\mu$ in this ansatz. Also the correction from bulk viscosity is naively expected to be smaller than that of shear viscosity, because $C_2 = -\frac{2}{5} C_1$ holds.

Thirdly, it is important to explicitly consider a multi-component system for estimations of the viscous corrections Eqs.~(\ref{eq:D0})-(\ref{eq:B2}) because the deviation of the distribution for the $i$-th particle species $\delta f_i$ generally depends on whether it is the only component or one of the components in a multi-component gas. This comes from the fact that $J_{mn}$'s, and thus the prefactors, include information of all particle species like thermodynamic variables. It should be noted that the difference becomes negligible if one treats the first order theory and assume that the shear viscosity $\eta$ is proportional to the entropy density $s$, because in the Boltzmann approximation we have
\begin{eqnarray}
\delta f^i_\mathrm{shear} &=& B_2 \times 2\eta \nabla _{\langle \mu} u _{\nu \rangle} \times p_i^\mu p_i^\nu \notag \\
&\propto & -\frac{1}{2J_{42}} \times 2s \times \nabla _{\langle \mu} u _{\nu \rangle} p_i^\mu p_i^\nu \notag \\
&\approx & -\frac{1}{2J_{42}} \times \frac{2J_{42}}{T^3} \times \nabla _{\langle \mu} u _{\nu \rangle} p_i^\mu p_i^\nu \notag \\
&=& C^i(T).
\label{eq:shear+1st}
\end{eqnarray}
Here we used the relation $J_{42} \approx sT^3$ in the Boltzmann approximation \cite{Israel:1979wp}. We see that $C^i(T)$ is a function dependent on the index $i$ and the temperature but independent of what other components are in the gas.

\section{The model}
\label{sec3}

\subsection{Equation of State and Transport Coefficients}

\begin{figure}[htb]
\includegraphics[width=3.4in]{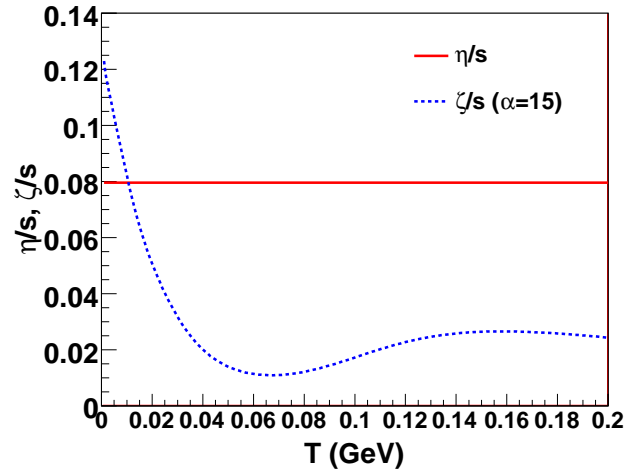}
\caption{(Color online) $\eta/s$ (solid) and $\zeta/s$ with $\alpha = 15$
(dotted) as functions of temperature in a resonance gas model.}
\label{fig:viscosity}
\end{figure}

Following the discussion in the previous section, we estimate the thermodynamic quantities and the prefactors appearing in $\delta f$. As the model of the equation of state, we consider the 16-component hadron resonance gas, which has mesons and baryons with mass up to $\Delta (1232)$. Hereafter we take the Landau frame and set the system to be baryon free. The models for the shear viscosity $\eta$ and the bulk viscosity $\zeta$ are taken from Refs. \cite{Son_visc} and \cite{Weinberg:1971mx}:
\begin{eqnarray}
\label{eq:eta}
\eta &=& \frac{1}{4\pi}s, \\
\label{eq:zeta}
\zeta &=& \alpha \bigg( \frac{1}{3}-c_s^2 \bigg)^2 \eta ,
\end{eqnarray}
where $c_s = \sqrt{\frac{\partial P}{\partial e_0}}$ is the sound velocity. The factor $\alpha$ in the bulk viscosity is set to 15 unless there is further notification.
Figure \ref{fig:viscosity} shows temperature dependence of 
$\eta/s$ and $\zeta/s$ for a hadronic resonance gas model.
In the temperature range relevant to relativistic heavy ion collisions
$0.1 \lsim T \lsim 0.2$ GeV, 
$\zeta/s$ is several factors smaller than $\eta/s$.
So one might expect that the effect of the bulk viscosity
is small. This is not necessarily true since the corrections in $\delta f$ 
appear as a combination of transport coefficients and the prefactors
discussed later.

%

Figure 2 shows the numerical results of the prefactors 
$D_0$, $B_0$, $\tilde{B}_0$, and $B_2$ appearing in $\delta f$
as functions of temperature $T$
in a hadronic resonance gas model.
%
Back to the discussion whether we should consider a non-zero trace tensor term or a trace part of a scalar term, we find that only the former is relevant for the 16-component hadron resonance gas case. This is because, if we choose the combination of the scalar term and the traceless tensor term, $D_0$, $B_0$, and $\tilde{B}_0$ diverge due to a change of the sign in the denominator (\ref{eq:detJ3}) at temperature below 0.2 GeV.
%
\begin{figure*}[htb]
 \includegraphics[width=3.4in]{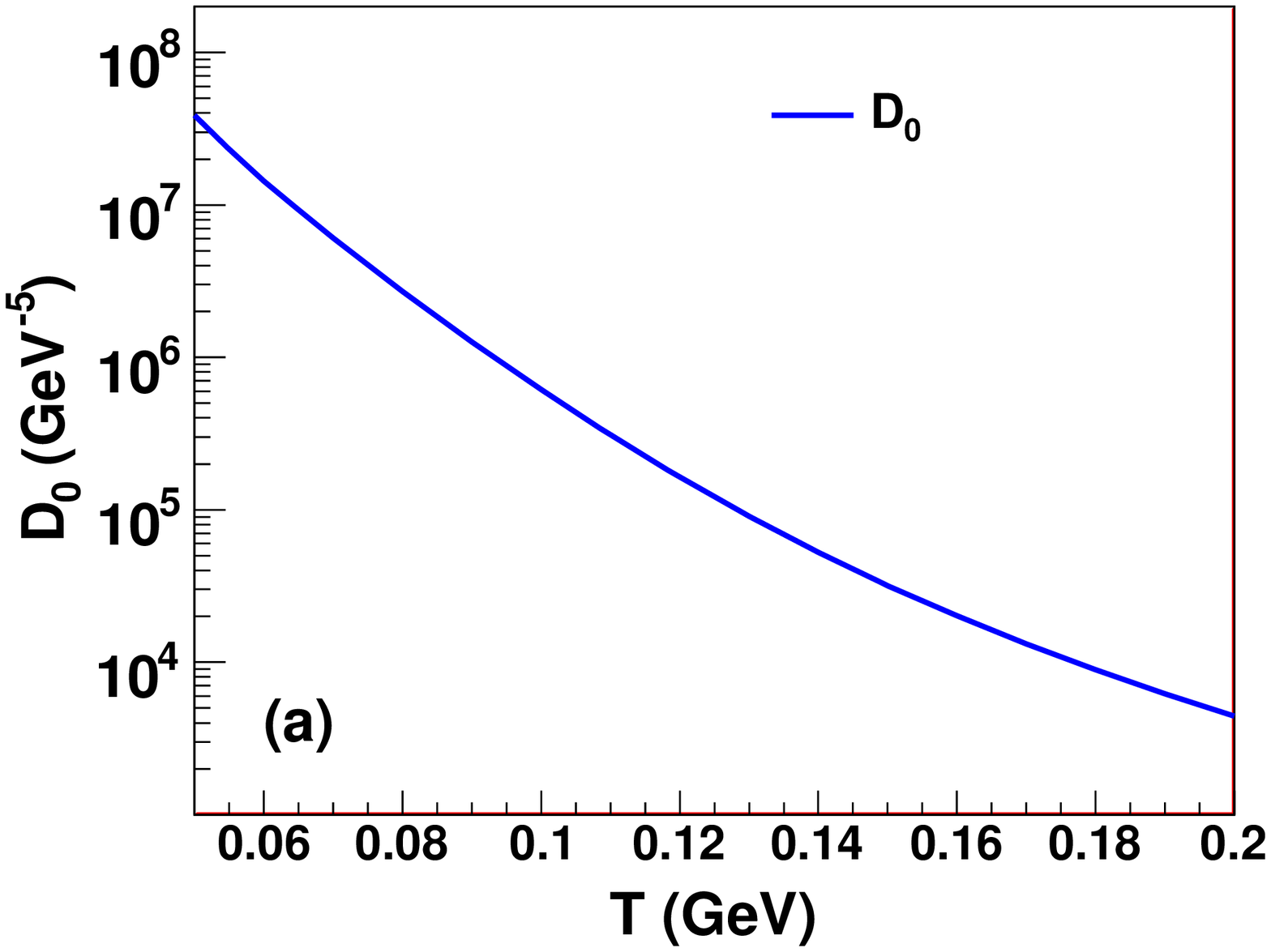}
 \includegraphics[width=3.4in]{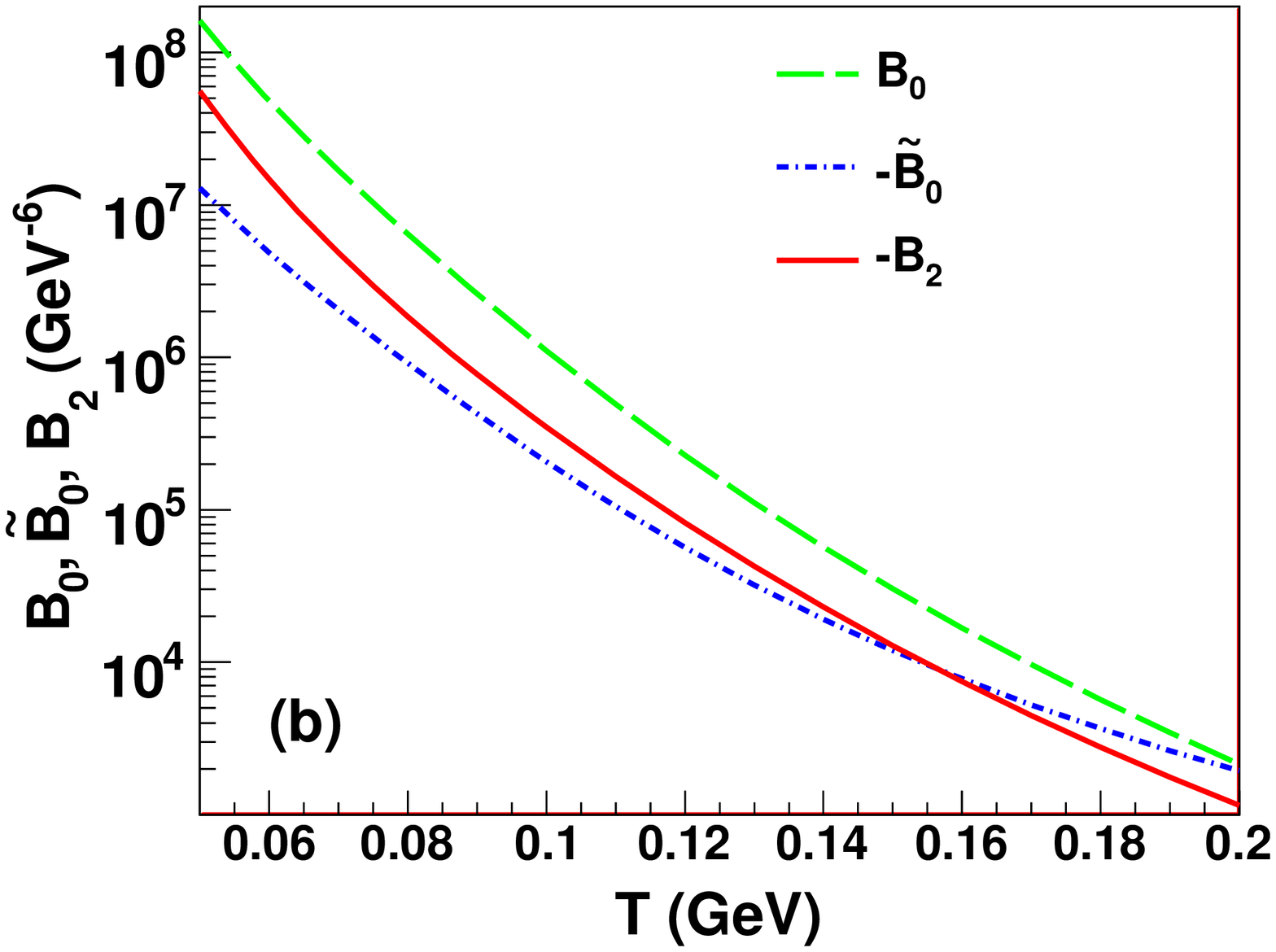}
 \caption{(Color online) (a) The prefactor in $\varepsilon _{\mu}$ for the bulk pressure, $D_0$, as a function of temperature. (b) The prefactors in $\varepsilon _{\mu \nu}$ for the bulk pressure, $B_0$ (dashed), and $\tilde{B}_0$ (dash-dotted), and that for the shear stress tensor $B_2$ (solid) as functions of temperature.
 }
 \label{fig:prefactors}
\end{figure*}
%

The numerical code to generate the prefactors in $\varepsilon _{\mu}$ and $\varepsilon _{\mu \nu}$ will become available \cite{webpageam}. One can set the components of a resonance gas and calculate the prefactors as well as any moment of $f_0 (1 + \epsilon f_0)$, $J_{mn}$.

\subsection{Particle Spectra and Flow}

In hydrodynamic analyses of the QGP created in relativistic heavy ion collisions, the Cooper-Frye formula is employed at freezeout. This converts the macroscopic variables into microscopic distributions, which enables us to compare the hydrodynamic results with experimental data. From a phenomenological point of view, it works as an interface from a hydrodynamic model to a cascade model. $p_T$-spectra of the $i$-th particle is given as, using the Cooper-Frye formula \cite{Cooper:1974mv},
\begin{equation}
\label{eq:ptspectra}
\frac{d^2N_i}{d^2p_Tdy} = \frac{g_i}{(2\pi )^3} \int _\Sigma p^\mu _i d \sigma _\mu (f^i_0 + \delta f^i),
\end{equation}
where $p_T$ and $y$ denote the transverse momentum and the rapidity. The equilibrium distribution $f_0$ is given as
\begin{equation}
\label{eq:f0}
f_0^i = \frac{1}{e^{(p^i \cdot u- b_i \mu _B )/T} - \epsilon}.
\end{equation}
The elliptic flow parameter is the coefficient of the second harmonics $\cos (2\phi_p )$ in a Fourier expansion of the azimuthal momentum distribution, where $\phi _p$ is the azimuthal angle in momentum space. It is defined as
\begin{equation}
\label{eq:v2pt}
v_2(p_T) = \frac{\int dy \int_0 ^{2\pi} d\phi_p \cos (2\phi_p ) \frac{dN_i}{d\phi_p p_Tdp_Tdy}}{\int dy \int_0 ^{2\pi} d\phi_p \frac{dN_i}{d\phi_p p_Tdp_Tdy}}.
\end{equation}
Viscous corrections are taken into account via (a) variation of the flow and (b) modification of the distribution function. Because we do not have a full (3+1)-dimensional viscous hydrodynamic flow, we focus on the latter in this study. Profiles of the flow $u^{\mu}$ and the freezeout hypersurface $d \sigma _\mu$, which are necessary for calculations of 
the Cooper-Frye formula (\ref{eq:ptspectra}),
are taken from (a) the Bjorken model \cite{Bjorken:1982qr}, (b) a blast wave model \cite{Teaney:2003kp}, and (c) a (3+1)-dimensional ideal hydrodynamic simulation \cite{Hiranov2eta}.

\section{Results}
\label{sec4}

We estimate the particle spectra and the elliptic flow coefficient of the negative pion with the mass $m = 0.139$ GeV. The freezeout temperature $T_f$ is set to $0.160$ GeV. This temperature is sufficiently near the QCD
(pseudo-)critical temperature. At $T_f = 0.160$ GeV, $\eta = 1.31\times 10^{-3}\ \mathrm{GeV}^3$ and $\zeta = 4.37\times 10^{-4}\ \mathrm{GeV}^3$ when $\alpha = 15$. In this study the Navier-Stokes limit is taken for the shear stress tensor and the bulk pressure, which means $\pi ^{\mu \nu} = 2\eta \nabla ^{\langle \mu} u ^{\nu \rangle} $ and $\Pi = -\zeta \partial _\mu u^\mu$ where $\nabla^\mu = \Delta^{\mu \nu} \partial _\nu$.

\subsection{Bjorken Flow}

We first consider the Bjorken model. We employ the expanding coordinates $(\tau, r, \phi, \eta _s)$, where proper time $\tau$, radius $r$, azimuthal angle $\phi$, and space-time rapidity $\eta _s$ are defined in the relations $t = \tau \cosh{\eta _s}$, $x = r \cos \phi$, $y = r \sin \phi$, and $z = \tau \sinh \eta _s$. In this frame, the Bjorken flow is written as
\begin{equation}
\label{eq:bjorken}
u^{\tau} = 1,\ u^r = u^\phi = u^{\eta _s} =0.
\end{equation}
The radius of the nuclei and the freezeout time are set to $R_0=10.0$ fm and $\tau=7.5$ fm, respectively. Elements of freezeout hypersurface only have the radial component:
\begin{equation}
d\sigma_{\tau} = \tau d \eta _s r dr d \phi ,\ d\sigma _r = d\sigma _\phi = d\sigma _{\eta _s} =0.
\label{eq:bjhypersurface}
\end{equation}

It is noteworthy that the Cooper-Frye formula for this model can be analytically expressed in the case of the Boltzmann approximation. See Appendix \ref{ar} for the details.

Figure \ref{fig:bjorken} shows the particle spectra with corrections from the bulk viscosity or the shear viscosity. The mean transverse momentum, $\langle p_T \rangle$, is lowered by the bulk viscosity in the case of the Bjorken flow. This can be interpreted from the sign of the shear stress tensor and the bulk pressure as follows. The bulk pressure works as a negative pressure, \textit{i.e.}, the pressure in the energy momentum tensor is effectively reduced to $P-| \Pi |$ because the system is expanding in the longitudinal direction and expansion scalar $\partial _\mu u^\mu = 1/\tau$ is positive. As a result, the number of particles with lower momenta increases. In a similar way, the shear viscosity is naively expected to enhance $\langle p_T \rangle$ in the mid-rapidity region because the pressure in the radial direction is increased by the decrease of the pressure in the longitudinal direction due to the fact that the shear stress tensor is traceless.

\begin{figure}[t]
\includegraphics[width=3.4in]{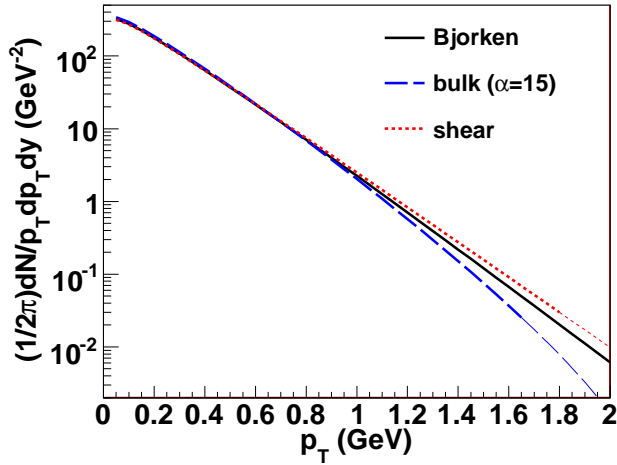}
\caption{(Color online) Viscous corrections to $p_T$-spectra with Bjorken flow.
Solid, thick dashed, and thick dotted lines
are, respectively, the results without any viscous corrections,
with the effect of bulk viscosity ($\alpha = 15$), and with the effect of the shear viscosity.
Thin dashed and dotted lines show that the absolute value of the ratio of the correction
to the ideal spectrum becomes greater than 0.5.
}
\label{fig:bjorken}
\end{figure}

\subsection{Flow from Blast Wave Model}

As a blast wave model, we follow Ref.~\cite{Teaney:2003kp}
\begin{eqnarray}
\label{eq:blastwave}
u^r &=& u_0\frac{r}{R_0}[1+u_2 \cos{(2\phi)}]\Theta (R_0 -r) 
\\
u^{\tau} &=& \sqrt{1+(u^r)^2} 
\\
u^\phi &=& u^{\eta _s} = 0
\end{eqnarray}
where $u_0=0.55$ and $u_2=0.2$. The radius of the nuclei and the freezeout time are set to $R_0=7.5$ fm and $\tau=5.25$ fm. The profile of the freezeout hypersurface is the same as Eq. (\ref{eq:bjhypersurface}) in the Bjorken model.

\begin{figure*}[htb]
\includegraphics[width=3.4in]{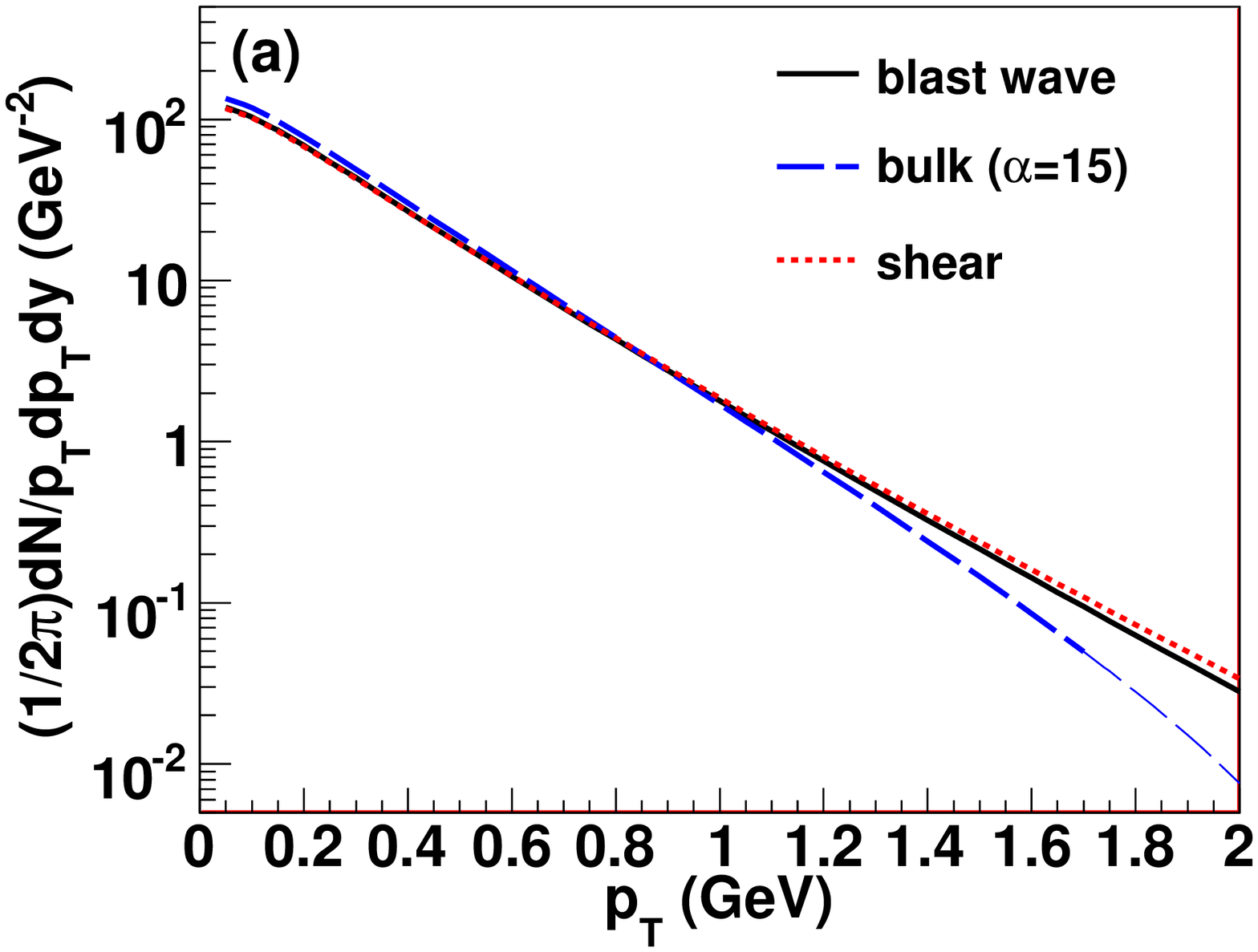}
\includegraphics[width=3.4in]{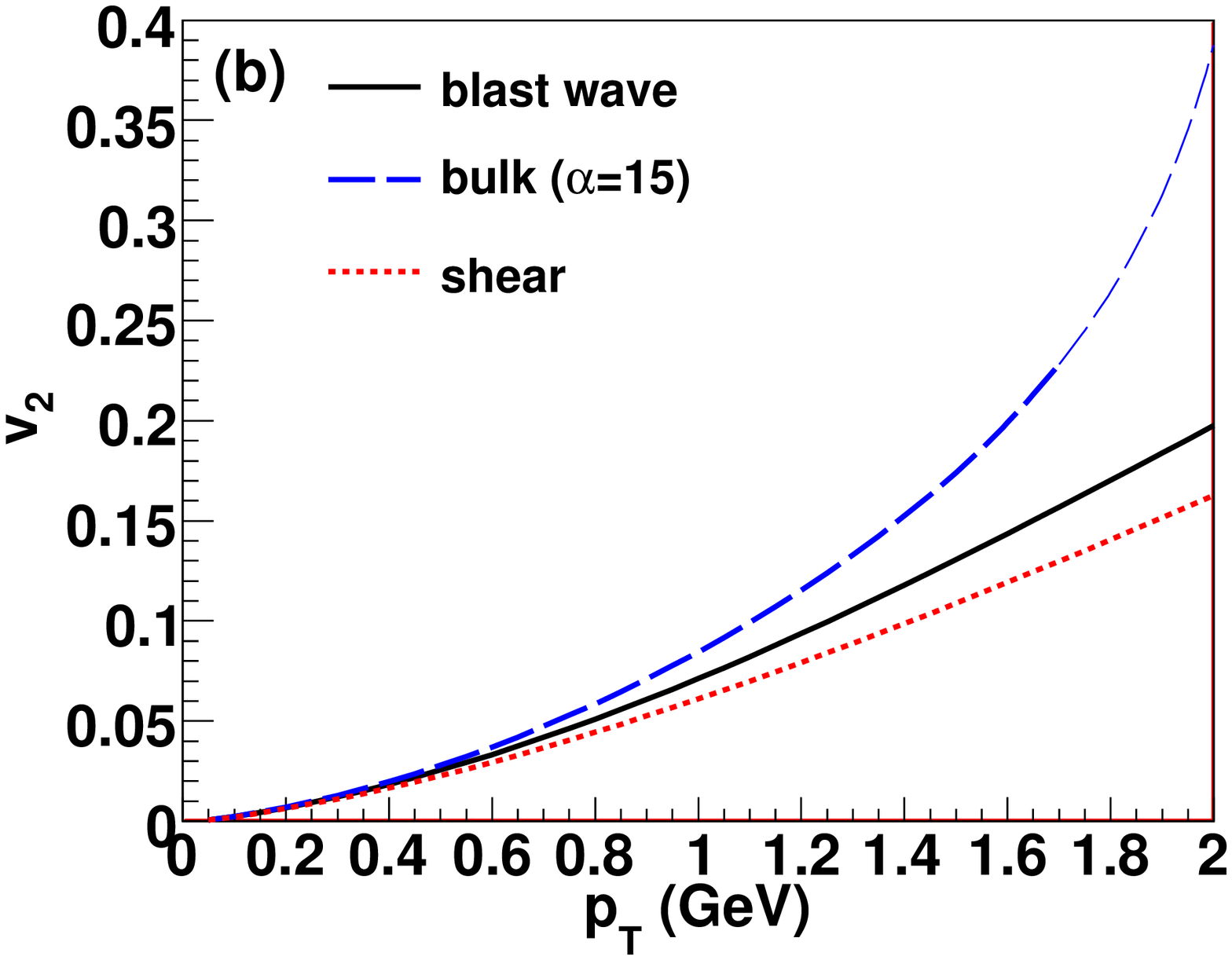}
\caption{(Color online) Viscous corrections to (a) $p_T$-spectra and (b) $v_2(p_T)$ in the blast wave model.
Solid, thick dashed, and thick dotted lines
are, respectively, the results without any viscous corrections,
with the effect of bulk viscosity ($\alpha = 15$), and with the effect of shear viscosity.
Thin dashed line shows that the absolute value of the ratio of the correction 
to the ideal spectrum becomes larger than 0.5.
}
\label{fig:blastwave}
\end{figure*}

Again the bulk viscosity lowers the spectrum as shown in Fig.~\ref{fig:blastwave}. The elliptic flow parameter $v_2(p_T)$, on the other hand, is \textit{enhanced} by the bulk viscosity. The counter-intuitive result can be interpreted as a fact that the slope of differential $v_2(p_T)$ can be given as $dv_2(p_T)/dp_T \simeq v_2/\langle p_T \rangle$ \cite{HG05}. The average $v_2$ is not much affected by the viscous correction in the case of the blast wave model. Effects of the shear viscosity can be explained likewise.\\

\subsection{(3+1)-Dimensional Ideal Hydrodynamic Flow}

In the last example, we demonstrate the effect of the bulk viscosity
by employing flow profiles from a full (3+1)-dimensional ideal hydrodynamic
simulation in Au+Au collisions at $\sqrt{s_{NN}} = 200$ GeV.
Impact parameter is taken to be 7.2 fm.
For details of this particular hydrodynamic model, see Refs. \cite{HHKLN, Hiranov2eta}.
Here the hadronic equation of state in this hydrodynamic
model is exactly the same as the one considered here, \textit{i.e.},
a hadronic resonance gas model up to the mass of $\Delta(1232)$.
Note that these flow profiles together with temperature distributions
are publicly available \cite{webpage}. 
It should be also noted that we have checked that $\Pi/P$ and $\pi^{\mu \nu}/T_0^{\mu \nu}$
at each freezeout position
have moderate values, at most of the order of unity.

\begin{figure*}[htb]
\includegraphics[width=3.4in]{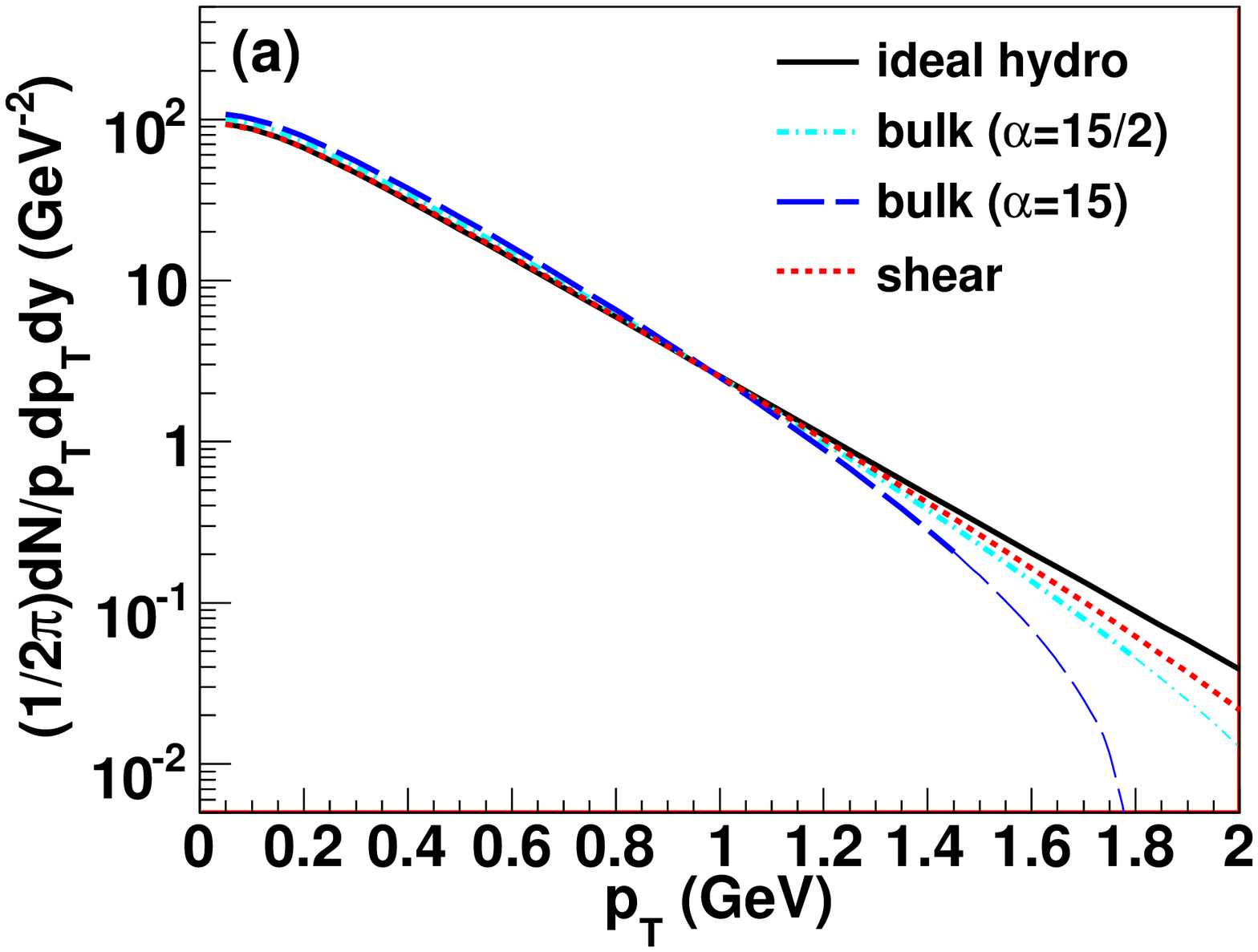}
\includegraphics[width=3.4in]{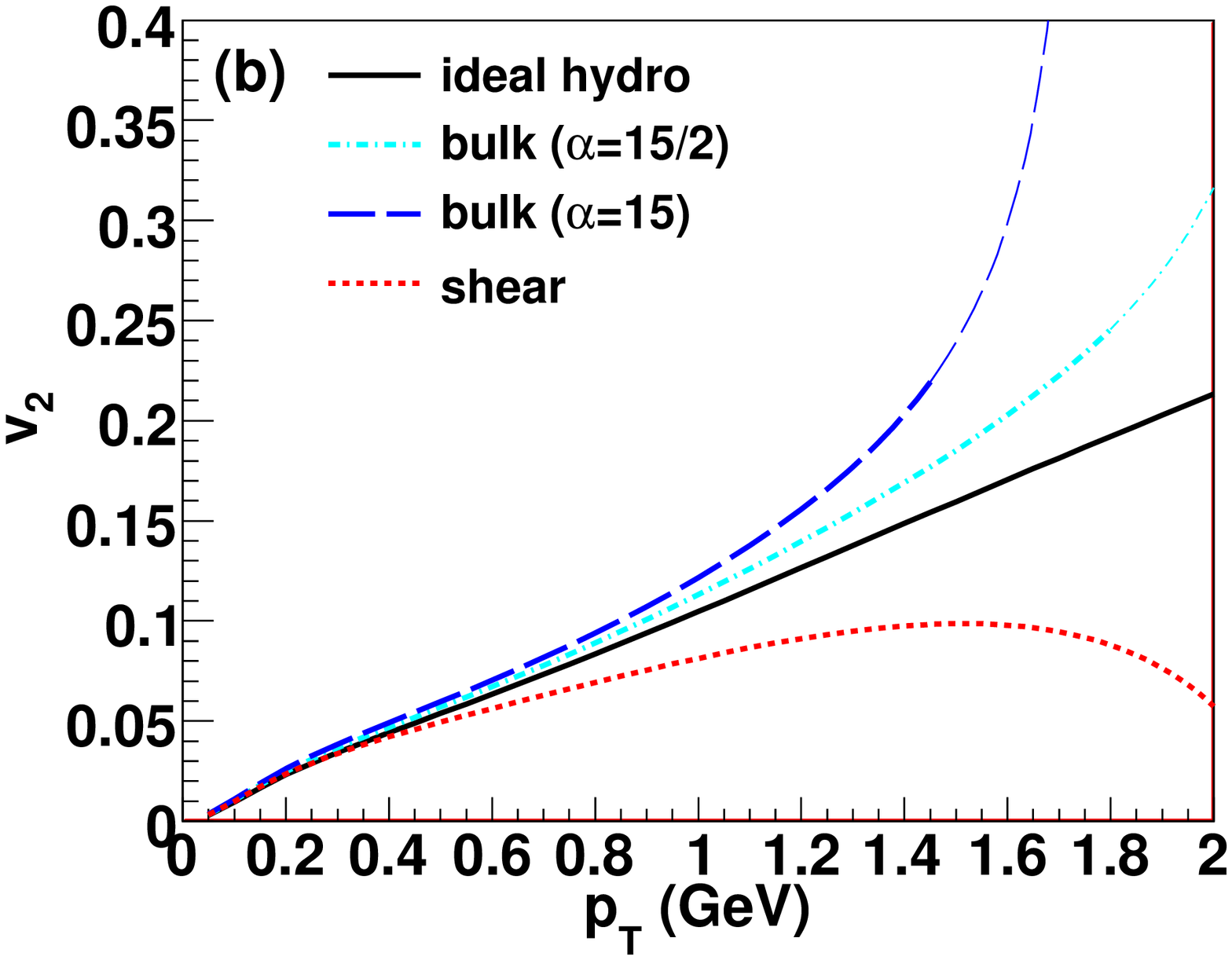}
\caption{(Color online) Viscous corrections to (a) $p_T$-spectra and (b) $v_2(p_T)$ with ideal hydrodynamic flow.
Solid, thick dash-dotted, thick dashed, and thick dotted lines
are, respectively, the results without any viscous corrections,
with the effect of bulk viscosity with $\alpha = 15/2$ and with $\alpha = 15$, and with the effect of shear viscosity.
Thin dash-dotted and dashed lines show that the absolute value of the ratio of the correction 
to the ideal spectrum becomes larger than 0.5.
}
\label{fig:idealhydro}
\end{figure*}

We see in Fig.~\ref{fig:idealhydro} that \textit{both} shear and bulk effects lower the spectrum in the high $p_T$ region. The enhancement of $v_2(p_T)$ by the bulk viscosity is due to the decrease in mean $p_T$ of the spectrum as discussed in the previous subsection. The non-triviality lies in the correction of the shear viscosity: It lowers the $p_T$-spectrum and still suppresses $v_2(p_T)$. This behavior was also observed for a (2+1)-dimensional viscous hydrodynamic flow \cite{Song:2007ux}.

The viscous corrections
in these calculations may have been overestimated for two reasons. Firstly, we neglected $\delta u^{\mu} = u^{\mu} - u^{\mu}_0$ by considering the ideal hydrodynamic flow. Viscosity tends to make the system moderate and consequently
thermodynamic forces, in particular, derivatives of the flow are
expected to become smaller in the case of viscous hydrodynamic flow.
This would reduce the amount of corrections considered here.
Secondly, the bulk pressure and the shear stress tensor are estimated in the first order theory, which naively means that
they correspond to asymptotic values
at the time much longer than relaxation times in the second order theory. 

%

\begin{figure*}[htb]
\includegraphics[width=3.4in]{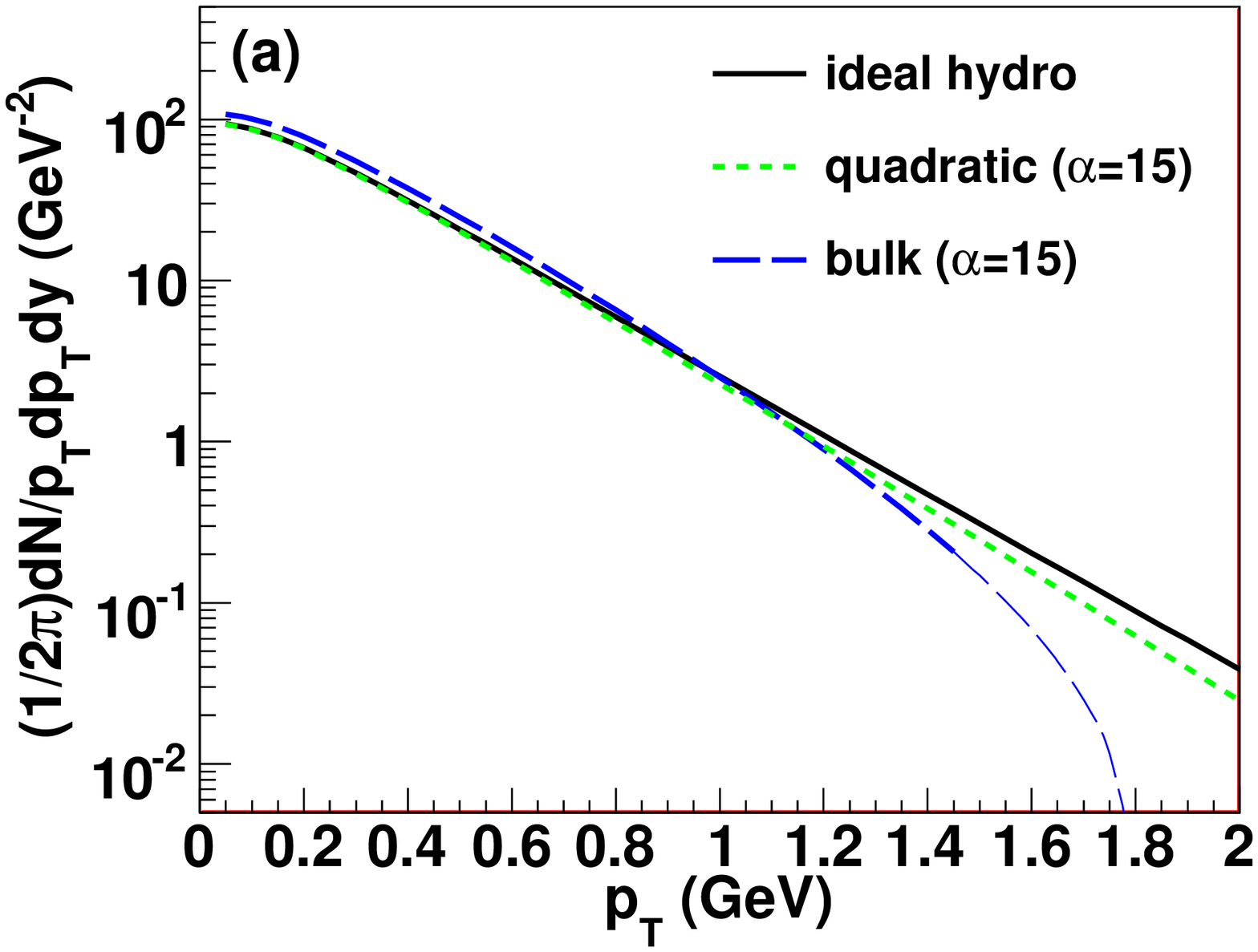}
\includegraphics[width=3.4in]{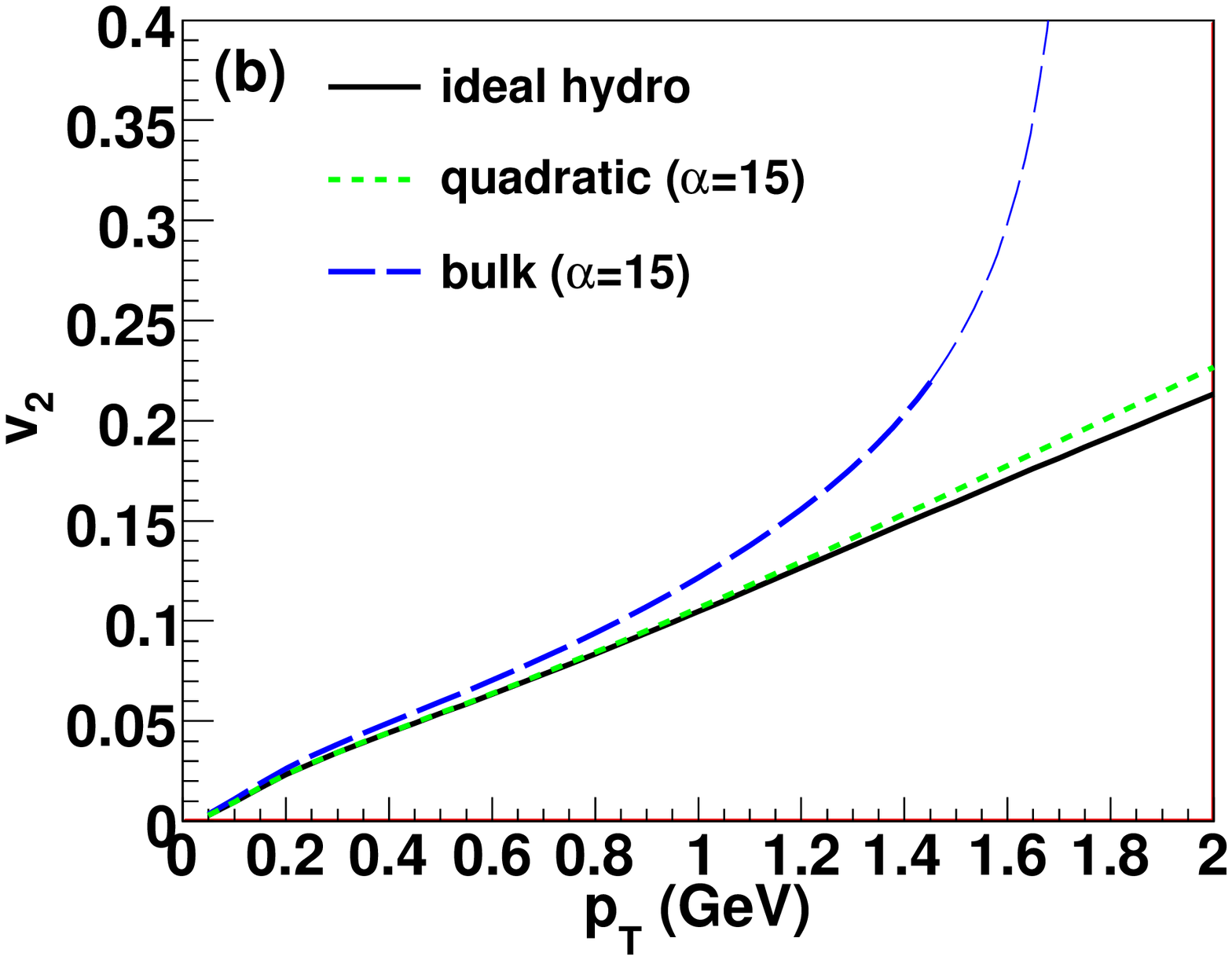}
\caption{(Color online) Viscous corrections to (a) $p_T$-spectra and (b) $v_2(p_T)$ with (thick dotted)
or without (thick dashed) the quadratic ansatz.
Solid line is the result from an ideal hydrodynamic model.
Thin dashed line shows that the absolute value of the ratio of the correction 
to the ideal spectrum becomes larger than 0.5.
}
\label{fig:quadratic}
\end{figure*}

Figure \ref{fig:quadratic} shows comparison of $p_T$-spectra
and $v_2(p_T)$ between
with and without the quadratic ansatz.
Neither the amount of the correction nor its $p_T$ dependence
seems to be similar to each other.
The quadratic ansatz greatly underestimates the effects
of the bulk viscosity.
Therefore, the proper
procedure to obtain
the prefactors in $\delta f$ should be made as discussed in Sec.~\ref{sec2}
to correctly calculate the distortion of particle spectra
due to the bulk viscosity. 

Finally when \textit{both} shear viscosity and bulk viscosity are considered, the slope of the particle spectra becomes steeper but that of $v_2(p_T)$ becomes moderate compared with that of the ideal distribution. 
The reason for the latter would be that the effect of the bulk viscosity and that of the shear viscosity 
accidentally cancels each other in the low $p_T$ region.

\begin{figure*}[htb]
\includegraphics[width=3.4in]{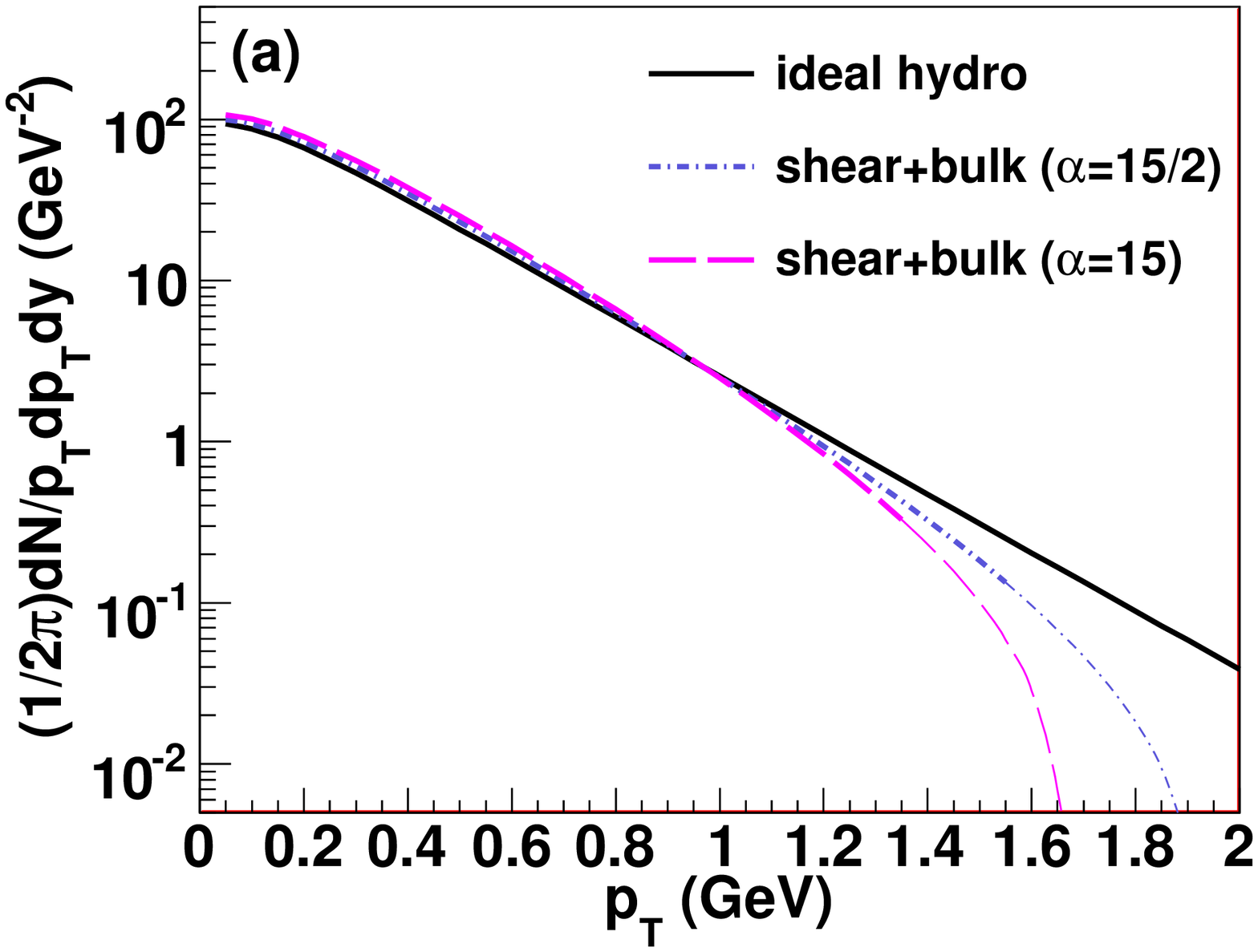}
\includegraphics[width=3.4in]{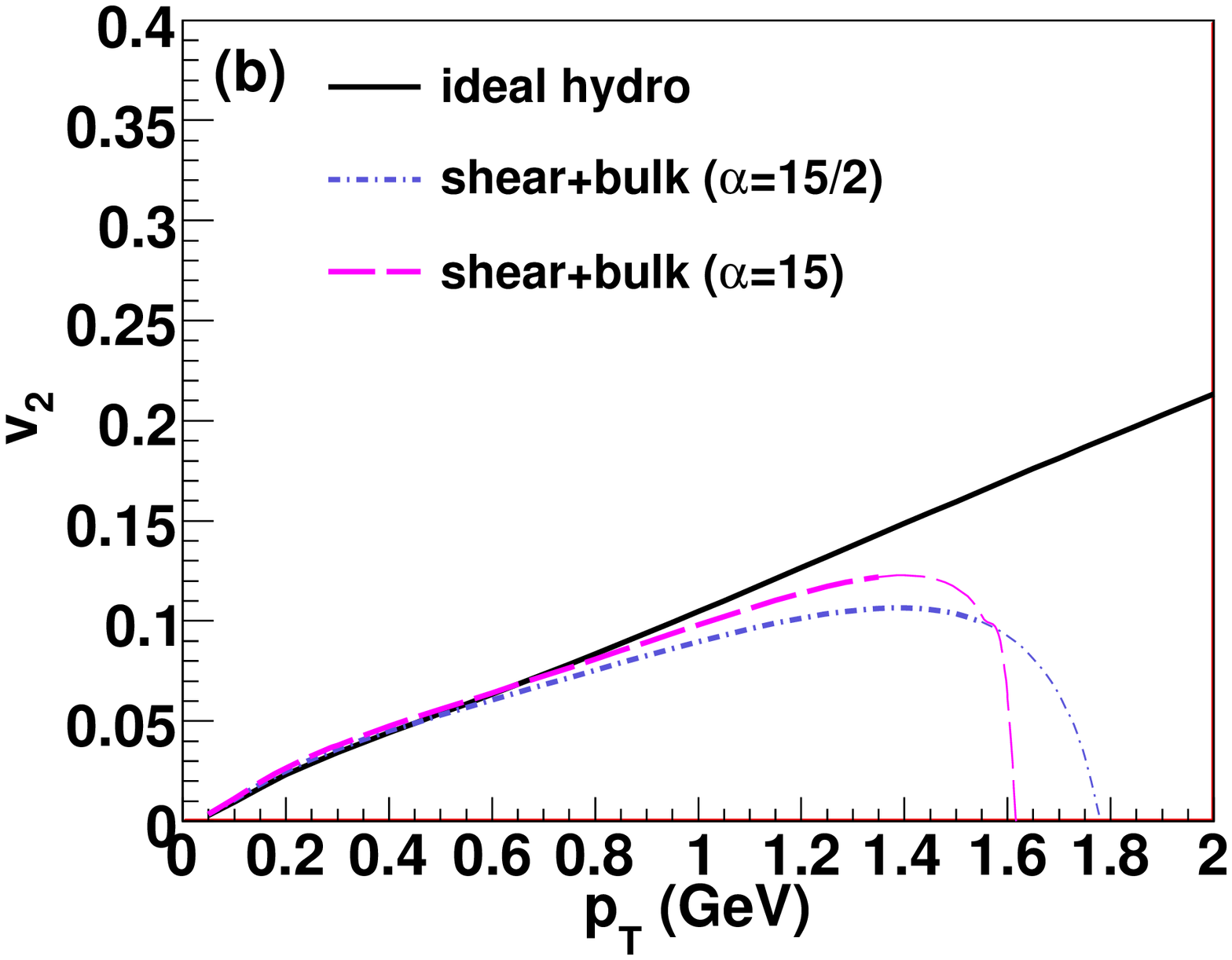}
\caption{(Color online) Viscous corrections of \textit{both} shear viscosity and bulk viscosity to (a) $p_T$-spectra and (b) $v_2(p_T)$.
Solid, thick dash-dotted, and thick dashed lines
are, respectively, the results without any viscous corrections,
with the effect of both shear and bulk viscosities with $\alpha = 15/2$, and with  $\alpha = 15$.
Thin dash-dotted and dashed lines show that the absolute value of the ratio of the correction 
to the ideal spectrum becomes larger than 0.5.
}
\label{fig:both}
\end{figure*}

\section{Conclusions}
\label{sec5}

We estimated the viscous corrections on the phase space distribution of a relativistic gas in a multi-component system. We found that generalization to a multi-component gas involves some subtleties. Firstly, the trace part of the tensor term and the scalar term in the distortion of the distribution function are equivalent for a single component gas, but not for a multi-component one. Numerical calculations suggest that one should have the trace part in the case of the 16-component hadron resonance gas.
Secondly, if one take zero net baryon density limit, one also has to take the limit for the Landau matching condition for the baryon number current because it does yield a finite relation. Since the number of equations does not change, we can uniquely determine the prefactors for the bulk pressure appearing in the distortion of the distribution. It is not desirable to introduce an additional ansatz that violates the matching conditions, since such an ansatz makes the system thermodynamically unstable in the first order theory.
Thirdly, the deviation of the distribution for the $i$-th gas component $\delta f_i$ is generally different depending on whether it is the only component or one of the components. This comes from the fact that the prefactors in $\varepsilon _{\mu}$ and $\varepsilon _{\mu \nu}$, like thermodynamic variables, include information of all the components in the gas. 

For hydrodynamic models of the QGP created in relativistic heavy ion collisions, the Cooper-Frye formula is necessary to convert macroscopic variables into microscopic distribution at freezeout. This enables us to compare the hydrodynamic results with experimental data, or to see further development of the hadronic matter in a cascade model. Non-equilibrium effects are taken into account in the formula via the variation of the flow and the modification of the distribution. We focused only on the latter and estimated the viscous corrections on the $p_T$-spectra and the elliptic flow $v_2(p_T)$ following the aforementioned discussion. Profiles of the flow and the freezeout hypersurface are taken from the Bjorken model with cylindrical geometry, the blast wave model with azimuthally anisotropic flow, and a (3+1)-dimensional ideal hydrodynamic model for the numerical calculations. We found that corrections of the bulk viscosity due to the distortion of the distribution have a visible effect on particle spectra and elliptic flow coefficient $v_2(p_T)$. This implies the importance of the bulk viscosity in constraining the transport coefficients with better accuracy from experimental data.

Quantitatively speaking, the viscous effects might have been overestimated because we approximated the shear stress tensor and the bulk pressure
as the ones in the first order theory and unlike in the second order theory, no relaxation effects are considered. As for the case of the estimation with an ideal hydrodynamic flow, the viscous corrections might be further exaggerated because the derivatives of the ideal hydrodynamic flow are generally larger than those of real viscous hydrodynamic flow. Considering the fact that the shear viscosity also has non-trivial effects on particle spectra depending on the flow profile, a full (3+1)-dimensional viscous hydrodynamic flow is necessary to see more realistic behavior of $p_T$-spectra and $v_2 (p_T)$.

\acknowledgments
The authors acknowledge fruitful discussions with T.~Hatsuda, T.~Kunihiro, and S.~Muroya.
The authors are also grateful for valuable comments by G.~S.~Denicol.
The work of T.H. was partly supported by
Grant-in-Aid for Scientific Research
No.~19740130 and by Sumitomo Foundation
No.~080734.  

\appendix

\section{The Landau Matching Conditions}
\label{lmc}

The Landau matching conditions ensure the thermodynamic stability in the first order theory. If the entropy current $s^\mu$ had a term proportional to $\Pi u^\mu$ in the non-equilibrium case, the derivative $\partial (u_\mu s^\mu) / \partial \Pi |_{\Pi=0} $ would be finite. This means that the system is not in a maximum entropy state, \textit{i.e.}, is not thermodynamically stable. The matching conditions take out such unwanted
terms.
One can explicitly show by inserting phase space distribution
into the definition of the entropy current in the relativistic kinetic theory
\begin{eqnarray}
s^\mu &=& - \sum _i \int \frac{g_i d^3p}{(2\pi )^3 E}p^\mu [(1+\epsilon f^i)\log{(1+\epsilon f^i)} + f^i \log{f^i}] \notag \\
&=& s u^\mu + \sum _i \int \frac{g_i d^3p}{(2\pi )^3 E}p^\mu \delta f_i \frac{p^\beta u_\beta - b_i \mu _B}{T} + \mathcal{O}[(\delta f _i )^2] \notag \\
&\approx& s u^\mu + \frac{u^\mu u_\alpha + \Delta^\mu _{\ \alpha}}{T} u_\beta \sum _i \int \frac{g_i d^3p}{(2\pi )^3 E}p^\alpha p^\beta \delta f_i \notag \\
&-& \mu _B \frac{u^\mu u_\alpha + \Delta^\mu _{\ \alpha}}{T} \sum _i \int \frac{b_i g_i d^3p}{(2\pi )^3 E}p^\alpha \delta f_i 
\label{eq:maxs}
\end{eqnarray}
\noindent
Here we neglect higher order terms 
in the third line for simplicity. By using the Landau matching conditions,
one can eliminate correction terms proportional to $u^\mu$
\begin{eqnarray}
s^\mu &=& s u^\mu + \frac{u^\mu}{T} u_\alpha \delta T^{\alpha \beta} u_\beta + \frac{1}{T} \Delta^\mu _{\ \alpha} u_\beta \delta T^{\alpha \beta} \notag \\
&-& \mu _B \frac{u^\mu}{T} u_\alpha \delta N_B^\alpha - \mu _B \frac{1}{T} \Delta^\mu _{\ \alpha} \delta N_B^\alpha \notag \\
&=& s u^\mu + \frac{W^\mu - \mu _B V^\mu}{T} 
\label{eq:maxs2}
\end{eqnarray}
\noindent
This expression 
includes only non-equilibrium corrections at the first order 
with respect to dissipative
currents which
are perpendicular to $u^\mu$.
This is exactly the generalization of the entropy current
in ideal hydrodynamics
to the one in viscous hydrodynamics
as discussed in Ref. \cite{Israel:1979wp}:
\begin{eqnarray}
s^\mu & =& \frac{s}{n_B} N_B^\mu+ \frac{q^\mu}{T} \notag \\
&=& \frac{P u^\mu + T^{\mu \nu} u_\nu - \mu _B N^\mu _B}{T}.
\label{eq:maxs3}
\end{eqnarray}

\section{Analytic Expressions of Spectra}
\label{ar}
We can analytically express the corrections of both shear viscosity and bulk viscosity on the Copper-Frye formula for the Bjorken model in the Boltzmann approximation following Ref.~\cite{Teaney:2003kp}. 
The flow and the freezeout hypersurface of the Bjorken scaling solution
with cylindrical geometry are
\begin{equation}
u^{\tau} = 1,\ u^r = u^\phi = u^{\eta _s} =0,
\label{eq:bjflow2}
\end{equation}
and
\begin{equation}
d\sigma_{\tau} = \tau d \eta _s r dr d \phi ,\ d\sigma _r = d\sigma _\phi = d\sigma _{\eta _s} =0.
\label{eq:bjhypersurface2}
\end{equation}
Since the momentum $p^\tau$ is given by $m_T \cosh{(y-\eta_s)}$ where $m_T$ is the transverse mass, the particle spectra is written as
\begin{eqnarray}
\frac{d^2N}{d^2p_Tdy} &=& \frac{g}{(2\pi )^3} \int _0 ^{R_0} r dr \int _0 ^{2\pi} d\phi \int _{-\infty} ^\infty \tau d\eta_s \notag \\
&\times& m_T \cosh{(y-\eta_s)} f.
\label{eq:bjcf}
\end{eqnarray}
Here we dropped the index $i$ of particle species for simplicity. The viscous corrections is now analytically expressed as
\begin{eqnarray}
\frac{d^2 \delta N_\mathrm{bulk}}{d^2p_Tdy} &=& \frac{g}{(2\pi )^3} \pi R_0^2 \cdot \tau m_T 2K_1(x) \notag \\
&\times & \frac{\zeta}{\tau} \bigg[ B_0 m^2 + \frac{D_0}{2} m_T  \frac{K_2(x)+1}{K_1(x)} \notag \\
&& \ \ + \frac{\tilde{B}_0 - B_0}{4} m_T^2 \bigg( \frac{K_3(x)}{K_1(x)} + 3 \bigg) \bigg] ,
\label{eq:bjcfb}
\end{eqnarray}
and
\begin{eqnarray}
\frac{d^2 \delta N_\mathrm{shear}}{d^2p_Tdy} &=& \frac{g}{(2\pi )^3} \pi R_0^2 \cdot \tau m_T 2K_1(x) \notag \\
&\times& \frac{2\eta}{3\tau} B_2 \bigg[ \frac{m_T^2}{2}  \bigg( \frac{K_3(x)}{K_1(x)} - 1 \bigg) - p_T^2 \bigg] .
\label{eq:bjcfs}
\end{eqnarray}
where $x = \frac{m_T}{T}$ and the modified Bessel function is defined as
\begin{equation}
K_n (x) = \int _0 ^\infty e^{-x\cosh{(t)}} \cosh{(nt)} dt.
\label{eq:modbessel}
\end{equation}
The bulk pressure and the shear stress tensor are estimated in the first order theory.

%

\end{document}